\documentclass[aps,pra,twocolumn,superscriptaddress]{revtex4-1}

\usepackage{amsmath}
\usepackage{cases}
\usepackage{graphicx}
\usepackage{mathtools} 
\usepackage[T1]{fontenc} 
\usepackage[utf8]{inputenc} 
\usepackage{amsfonts} 
\usepackage{bm}

\usepackage{enumitem}
\usepackage{multirow}
\usepackage{tabularx}
\usepackage{longtable}
\usepackage{float}

\usepackage{hyperref}
\usepackage[capitalise,nameinlink]{cleveref}
\usepackage[usenames,dvipsnames]{xcolor}
\hypersetup{
colorlinks=true 
linkcolor=Maroon, 
citecolor=OliveGreen, 
filecolor=magenta, 
urlcolor=Blue   
}
\crefname{ansatz}{Ansatz.}{Ansatzes.}

\newcommand{\PNNL}{%
    \affiliation{%
        Physical \& Computational Science Division, 
        Pacific Northwest National Laboratory, 
        Richland WA 99352, USA}}

\begin{document}
\title{Integrating Subsystem Embedding Subalgebras and Coupled Cluster Green's Function: A Theoretical Foundation for Quantum Embedding in Excitation Manifold}
\author{Bo Peng} 
\email{peng398@pnnl.gov}
\PNNL
\author{Karol Kowalski} 
\PNNL

\begin{abstract}
In this study, we introduce a novel approach to coupled-cluster Green's function (CCGF) embedding by seamlessly integrating conventional CCGF theory with the state-of-the-art sub-system embedding sub-algebras coupled cluster (SES-CC) formalism. This integration focuses primarily on delineating the characteristics of the sub-system and the corresponding segments of the Green's function, defined explicitly by active orbitals.
Crucially, our work involves the adaptation of the SES-CC paradigm, addressing the left eigenvalue problem through a distinct form of Hamiltonian similarity transformation. This advancement not only facilitates a comprehensive representation of the interaction between the embedded sub-system and its surrounding environment but also paves the way for the quantum mechanical description of multiple embedded domains, particularly by employing the emergent quantum flow algorithms.
Our theoretical underpinnings further set the stage for a generalization to multiple embedded sub-systems. This expansion holds significant promise for the exploration and application of non-equilibrium quantum systems, enhancing the understanding of system-environment interactions. In doing so, the research underscores the potential of SES-CC embedding within the realm of quantum computations and multi-scale simulations, promising a good balance between accuracy and computational efficiency.
\end{abstract}

\maketitle

\section{Introduction}

Quantum molecular systems situated in complex environments—often referred to as open quantum molecular systems—engage in dynamic exchanges with their surroundings. These exchanges, encompassing charge, spin, particle, and energy, are fundamental processes critical for a diverse range of scientific fields. These fields include but are not limited to non-equilibrium chemical transformations, transient spectroscopies, surface physics, and chemistry in catalysis, photo-induced charge transfer in molecular optoelectronics, and photo- and thermally-induced spin transitions in molecular spintronics. Moreover, understanding decoherence in molecular qubits and noise measurements in quantum thermodynamics is imperative for advancements in quantum information science. Accurately capturing the dynamical processes of these systems opens new avenues for the design and characterization of intricate quantum molecular systems, with far-reaching implications across these disciplines \cite{breuer2002theory, weiss2012quantum, may2011charge, wasielewski2020exploiting}.

Despite significant enhancements in computational capabilities over recent decades, the computational expense associated with describing complex molecular systems accurately remains a considerable challenge. Strategies to mitigate these computational demands, while preserving the integrity of high-fidelity descriptions, range from scaling and extrapolation methods to quantum embedding (QE) techniques \cite{wesolowski2015frozen,sun2016quantum}. Traditional scaling and extrapolation methods, while valuable, can prove prohibitively expensive or even infeasible when applied to systems with extensive degrees of freedom. In contrast, QE techniques have emerged as particularly potent tools, demonstrating remarkable efficacy in a variety of real-world scenarios at the cutting edge of chemical research, including areas like catalysis, biochemistry, and spectroscopy.

The development of an effective QE method generally begins with the partitioning of the targeted system into smaller, more manageable subsystems. This often involves defining an embedded region within a broader embedding environment. The method then strategically combines advanced, yet computationally intensive, treatments for the embedded region with more efficient computational approaches for the environmental degrees of freedom. This synergy aims to minimize the overall computational burden, making the comprehensive analysis more practical than if one were to attempt a singular, high-level computation for the entire system. This principle has guided the development of numerous QE methodologies, each with unique attributes tailored to specific types of system partitioning and inter-system interactions \cite{jones2020embedding,sushko2018embedding,libisch2014embedded,seijo1999abinitio,jacob2014subsystem,wesolowski2015frozen,sun2016quantum}.

Recognizing the importance of geometrically sensible system partitioning (e.g., along chemical bonds or between molecular clusters) and accurate handling of subsystem interactions, especially in scenarios like charge delocalization and quantum transport, is essential. In these instances, interactions are so pivotal that quantum processes span regions significantly larger than the predefined embedded area. Accordingly, the detailed understanding and treatment of these interactions become paramount in the development of reliable and predictive models.

Current QE strategies can broadly classify into two main categories: those combining quantum mechanics with classical molecular mechanics (QM:MM) and those integrating various levels of quantum mechanical theories (QM:QM) \cite{Warshel1976theoretical,brunk2015mixed,svensson1996oniom,chung2015oniom}. The QM:QM category encompasses several sophisticated methods, including but not limited to, fully quantum mechanical treatment through \textcolor{black}{density functional theory (DFT)} embedding, partition DFT, potential-functional embedding, embedded mean-field theory (EMFT), Green’s function embedding, self-energy embedding, density matrix embedding theory (DMET), and stochastic embedding DFT \cite{jacob2014subsystem,govind1998accurate,huang2011quantum,manby2012simple,elliott2010partition,nafziger2014density,huang2011potential,fornace2015embedded,inglesfield1981embedding,chibani2016self,lan2015towards,rusakov2019self,knizia2013density,li2019stochastic}. These approaches leverage various representations of quantum mechanics—such as single-particle density, single-particle Green's function, and single-particle density matrix—for performing density functional embedding, Green's function embedding, and density matrix embedding, respectively \cite{sun2016quantum,jones2020embedding}.

Despite the diverse theories and broad application domains, two predominant challenges arise:
\begin{itemize}
\item Constructing a unified embedding formalism that integrates various quantum mechanical representations—encompassing wave function, reduced density matrix, Green’s function, self-energy, among others—rigorously, while minimizing the embedding scheme's dependence on system partitioning.
\item Describing the coupling between the embedded region and the surrounding environment in a systematically improvable way. This approach should seamlessly connect the non-interacting subsystem limit (NSL) and the fully entangled limit (FEL) of the open system, thereby guiding the balance between accuracy and computational efficiency for various applications. 
\end{itemize}

\textcolor{black}{In response to these challenges, and particularly for the condensed matter community, previous efforts have introduced the coupled cluster approach with single and double excitations as imputiry solver to compute the impurity Green's function in dynamical mean-field theory~\cite{PhysRevB.100.115154} or self-energy embedding theory~\cite{shee2019CC}. The advantage of this method is its provision of an impurity solver with polynomial scaling, in contrast to exact diagonalization, which exhibits exponential scaling.}
In this paper, we introduce a new quantum embedding methodology that innovatively applies coupled-cluster subsystem embedding sub-algebra techniques \cite{kowalskiCCsubalgebra} alongside extensive many-body coupled-cluster representations of wave functions \cite{arponen83_311} and many-body coupled-cluster Green’s function theory \cite{nooijen92_55,nooijen93_15,nooijen95_1681,KK:peng2021ccgf}. Our approach presents a unique method for the rigorous partitioning of the Green's function of open quantum systems, effectively distinguishing between the low and high energy regimes of the spectrum. This distinction significantly enhances the precision of electronic structure representations and quantum dynamics descriptions of open systems, surpassing the constraints found in traditional embedding algorithms. Moreover, this methodology is inherently adaptable for multiscale/multiphysics simulations, promising extensive applicability.

The remainder of the paper is structured into four detailed sections. Section \ref{SESCC} will succinctly revisit the Sub-system Embedding Sub-algebras (SES) and its intricate coupled-cluster formalism (SES-CC), highlighting notable properties. Section \ref{LambdaSES} delves into the generalization of SES-CC, employing the conventional bi-orthogonal CC formalism, focusing on $\Lambda$ parametrization and its exponential alternatives. In Section \ref{SESCCGF}, we dedicate our efforts to developing the SES-CC Green's function formalism, illustrating the systematic decomposition of conventional CCGF calculations into distinct SES-CCGF computations for the embedded component and its corresponding environment. Section \ref{sec:numerical} provides a practical demonstration of the SES-CC Green's function approach, utilizing a three-site single impurity Anderson model (SIAM) to elucidate the workflow of SES-CCGF, particularly demonstrating how embedded subspace calculations can be conducted with precision. It is crucial to note that the analyses within this study are premised on the exact wave function limit formulations of CC and CC Green's function theories. Consequently, any simplifications or approximations are thereby a reduction from the equations derived in this exact limit.

\section{Methodology}\label{sec:method}

\subsection{A brief review of sub-system embedding subalgebras coupled cluster (SES-CC) formalism and its properties}\label{SESCC}

In the single reference coupled cluster (CC) approach, the electronic ground-state wave function $|\Psi\rangle$ of an $N$-electron system is assumed to be parametrized through an exponential anstaz, $|\Psi\rangle = e^T |\Phi\rangle$, where $|\Phi\rangle$ a single reference. In this exponential ansatz, the exponent operator $T$ (i.e. the cluster operator) comprises excitation operators that produce excited configurations from the reference. In the language of second quantization, $T$ can be expressed as
\begin{equation}
T=\sum_{k=1}^{m} \frac{1}{(k!)^2} \sum_{\substack{i_1,\ldots,i_k ; \\ a_1, \ldots, a_k}} t^{i_1\ldots i_k}_{a_1\ldots a_k} a_{a_1}^\dagger\cdots a_{a_k}^\dagger a_{i_k} \cdots a_{i_1}~.
\label{T}
\end{equation}
Here, the indices $i_1,i_2,\ldots$ ($a_1,a_2,\ldots$) denote occupied (unoccupied) spin-orbitals in the reference $|\Phi\rangle$, $a_p^\dagger$ ($a_p$) is the standard creation (annihilation) operator that add (remove) electron at spin-orbital $p$, and $t^{i_1\ldots i_k}_{a_1\ldots a_k}$'s denote cluster amplitudes, $m$ is the excitation level ($\le N$) that defines the approximate approach within the CC framework (e.g. $m=2$ for CCSD, \cite{purvis82_1910}, $m=3$ for CCSDT, \cite{ccsdt_noga,ccsdt_noga_err,SCUSERIA1988382} $m=4$ for CCSDTQ ($m=4$), \cite{Kucharski1991,ccsdtq_nevin}). Once $T$ is defined, we can plug the exponential ansatz into the Schr{\"o}dinger equation to get the energy-dependent form of the CC equation 
\begin{align}
(P^{(N)}+Q^{(N)}) H e^T|\Phi\rangle &= E_{\text{CC}}^{(N)} (P^{(N)}+Q^{(N)}) e^T |\Phi\rangle, \label{Ecc}
\end{align}
where $H$ is the electronic Hamiltonian, $E_{\rm CC}^{(N)}$ is the coupled cluster energy, and $P^{(N)}$ and $Q^{(N)}$ are operators that project Eq. (\ref{Ecc}) onto the reference and excited configurations of the $N$-electron system, respectively,
\begin{align}
P^{(N)} &= | \Phi \rangle \langle \Phi |, \notag \\
Q^{(N)} &= \sum_{k=1}^{m}~\sum_{\substack{i_1<\ldots<i_k; \\ a_1<\ldots <a_k}}
 |\Phi_{i_1\ldots i_k}^{a_1 \ldots a_k}\rangle\langle \Phi_{i_1\ldots i_k}^{a_1 \ldots a_k}|. \label{Qop}
\end{align}

\begin{figure}
    \centering
    \includegraphics[width=\linewidth]{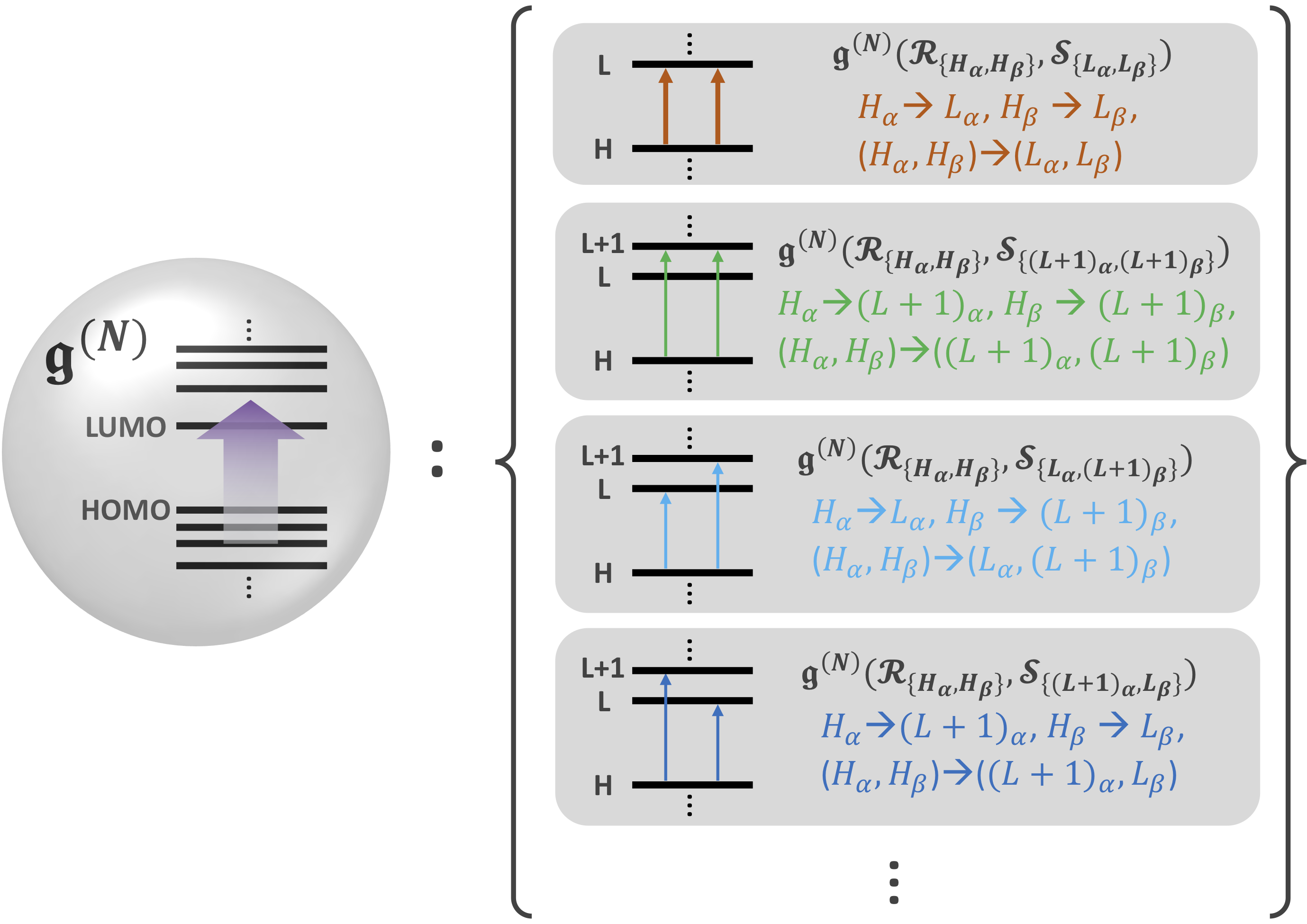}
    \caption{All the excitations in an $N$-electron system based on a reference $|\Phi\rangle$ form an excitation algebra $\mathfrak{g}^{N}$ that includes excitation subalgebras $\mathfrak{g}^{N}(\mathcal{R},\mathcal{S})$'s.
    \label{fig:ses}}
\end{figure}

Focusing on the excitation operators $T$ in Eq. (\ref{T}), as discussed in Ref. \citenum{kowalskiCCsubalgebra}, these operators constitute a commutative algebra, $\mathfrak{g}^{(N)}$, \textcolor{black}{(see Fig. \ref{fig:ses})}. Also, we can define a class of active space excitation/de-excitation subalgebras, denoted as $\mathfrak{h} \coloneqq \mathfrak{g}^{(N)}(\mathcal{R},\mathcal{S})$, which facilitate the excitation of electrons from an active subspace $\mathcal{R}$ within the occupied orbital space to an active subspace $\mathcal{S}$ within the virtual orbital space, encompassing all possible excitations. In the context of the particle-hole formalism, all subalgebras $\mathfrak{h}$ and the overarching excitation algebra $\mathfrak{g}^{(N)}$ are commutative. In particular scenarios where the subalgebras include either all occupied and selected virtual orbitals ($\mathcal{S}$) or a specific set of occupied orbitals ($\mathcal{R}$) and all virtual orbitals, we adopt the concise notations $\mathfrak{g}^{(N)}(\mathcal{S})$ and $\mathfrak{g}^{(N)}(\mathcal{R})$, respectively. In some instances, the notation $\mathfrak{g}^{(N)}(x_\mathcal{R},y_\mathcal{S})$ is employed to specify the number of orbitals, $x$ and $y$, included in the active subspaces $\mathcal{R}$ and $\mathcal{S}$, respectively.

The purpose of introducing the active-space excitation/de-excitation subalgebras lies in the fact that the cluster operator $T$ can be generally decomposed to the commutative components that belong to $\mathfrak{h}$ (denoted as the internal part, $T_{\rm int}(\mathfrak{h})$) and $\mathfrak{g}^{(N)}-\mathfrak{h}$ 
(denoted as an external part, $T_{\rm ext}(\mathfrak{h})$), respectively, i.e.
\begin{equation}
T=T_{\rm int}(\mathfrak{h})+T_{\rm ext}(\mathfrak{h}),~~\text{with}~~[T_{\rm int}(\mathfrak{h}), T_{\rm ext}(\mathfrak{h})]=0 .
\label{Tdecomp}
\end{equation}
Accordingly, the corresponding wave function can be rewritten as
\begin{align}
|\Psi\rangle &=  e^{T}|\Phi\rangle =
e^{T_{\rm ext}+ T_{\rm int}}|\Phi\rangle
= e^{T_{\rm ext}} |\Psi_{\rm int}\rangle.
\label{nildec_1}
\end{align}
where we denote $|\Psi_{\rm int}\rangle = e^{T_{\rm int}}|\Phi\rangle$ as the auxiliary \textcolor{black}{complete active space (CAS)} type wave function corresponding to the internal active subspace, and drop the explicit $\mathfrak{h}$-dependence in $T_{\rm int}$, $T_{\rm ext}$, and $|\Psi_{\rm int}\rangle$ for the simplification of notation (and same below).

The subalgebras $\mathfrak{h}$ can be further refined into one class that uniquely maps to the CC approximations if (\textbf{a}) $|\Psi(\mathfrak{h})\rangle$ preserves the symmetry properties of $|\Psi\rangle$ and $|\Phi\rangle$, and (\textbf{b}) the $e^{T_{\rm int}}|\Phi\rangle$ ansatz generates the full configuration interaction (FCI) expansion for the sub-system defined by the active subspace corresponding to the $\mathfrak{h}$ subalgebra. In Ref.  \citenum{kowalskiCCsubalgebra} this is referred as sub-system embedding subalgebras (SESs). 

The SES $\mathfrak{h}$ can be employed to derive an alternative representation of the CC equation that is naturally adapted to the sub-system embedding. Following the decomposition of $T$, we can further decompose the projection operator $Q^{(N)}$ as
\begin{align}
Q^{(N)} = Q_{\rm int}^{(N)} + Q_{\rm ext}^{(N)}, \label{Qdecomp}
\end{align}
where $Q_{\rm int}^{(N)}$ and $Q_{\rm ext}^{(N)}$ are constructed from the excited configurations generated by respectively acting $T_{\rm int}$ and $T_{\rm ext}$ on the reference. \textcolor{black}{Denote
\begin{align}
    \mathcal{P}_{\rm int}^{(N)} = P^{(N)} + Q_{\rm int}^{(N)}
\end{align}
(such that $\mathcal{P}_{\rm int}^{(N)} + Q_{\rm ext}^{(N)} = \mathbf{1}^{(N)}$)
} and define $\overline{H}_{\rm ext} = e^{-T_{\rm ext}}He^{T_{\rm ext}}$, we can then re-write CC energy equation as
\begin{align}
\mathcal{P}_{\rm int}^{(N)} \overline{H}_{\rm ext} |\Psi_{\rm int}\rangle &= E_{\rm CC}^{(N)} \mathcal{P}_{\rm int}^{(N)} |\Psi_{\rm int}\rangle,  \label{SESCC}
\end{align}
for the SES $\mathfrak{h}$ with $T_{\rm ext}$ being obtained from the projections of $(He^T)_{\rm C}|\Phi\rangle$ on the excited configurations that are not included in the SES $\mathfrak{h}$
\begin{align}
Q_{\rm ext}^{(N)} \overline{H}_{N} |\Phi\rangle &= 0, \label{SESCC_T}
\end{align}
where the normal ordered Hamiltonian is defined as
\begin{align}
\overline{H}_N = e^{-T}He^T - E_{\rm CC}^{(N)} = \overline{H} - E_{\rm CC}^{(N)}.
\end{align}
Because $Q_{\rm ext}^{\rm (N)} |\Psi_{\rm int}\rangle = 0$, \textcolor{black}{the identity between $\overline{H}_{\rm ext}$ and $|\Psi_{\rm int}\rangle$ in Eq. (\ref{SESCC}) can be equivalently replaced by $\mathcal{P}_{\rm int}^{(N)}$, and Eq. (\ref{SESCC}) can be equivalently rewritten as
\begin{align}
    \mathcal{P}_{\rm int}^{(N)} \overline{H}_{\rm ext} \mathcal{P}_{\rm int}^{(N)} |\Psi_{\rm int}\rangle &= E_{\rm CC}^{(N)} \mathcal{P}_{\rm int}^{(N)} |\Psi_{\rm int}\rangle.  \label{SESCC_eig}
\end{align}
with $\mathcal{P}_{\rm int}^{(N)}|\Psi_{\rm int}\rangle$ denoting the CAS-type wave function projected to the active space.}
As can be seen from Eq. (\ref{SESCC_eig}), $E_{\rm CC}^{(N)}$ and $|\Psi_{\rm int}\rangle$ (and hence $T_{\rm int}$) can now be obtained by directly diagonalizing $\overline{H}_{\rm ext}$ in the $(P^{(N)}+Q_{\rm int}^{(N)})$ space. The generated effective Hamiltonian, 
\begin{align}
\overline{H}_{\rm ext}^{\rm eff} = 
\mathcal{P}_{\rm int}^{(N)} \overline{H}_{\rm ext} \mathcal{P}_{\rm int}^{(N)}, \label{eff}
\end{align}
is $\mathfrak{h}$-dependent, and provides a unique and formally exact way to downfold the full electronic Hamiltonian to the active space associated with the SES $\mathfrak{h}$. Due to this feature, SES has been proposed to work with double or generalized unitary CC formalism to provide effective Hermitian Hamiltonians in the variational quantum approaches, as reported in recent studies  \cite{Bauman2019SESDUCC, Metcalf2020VQEDUCC, bauman2020variational}, showing the promising ability of capturing a large part of dynamical electron correlation usually missing in relatively smaller active space quantum simulations.

\subsection{Extension of SES formalism for $\Lambda$ formalism}\label{LambdaSES}

In contrast to the $T$ operator, the de-excitation $\Lambda$ operator involves all the de-excitation operators $a_{i_1}^{\dagger} \cdots a_{i_k}^\dagger a_{a_k} \cdots a_{a_1}$ defined with respect to the $|\Phi\rangle$ reference. Several CC formulations such as  linear response CC, analytical CC gradients, and CC Green's function formulations employ the so-called bi-orthogonal CC formalism,  \cite{arponen83_311,Szalay1995ECC, PIECUCH1999XCC,salter1987property,stanton93_7029,monkhorst77_421,jorgensen90_3333,nooijen92_55, nooijen93_15, nooijen95_1681,kowalski14_094102,kowalski16_144101,kowalski16_062512,kowalski18_4335,helgaker2014molecular, Schirmer2010BiCC} which utilizes different parametrization for the  bra  ($\langle\Psi|$) of the ground-state wave functions 
\begin{align}
\langle\Psi|   &= \langle\Phi|(1+\Lambda) e^{-T} ~~, \label{bra1} 
\end{align}
where it is easy to verify that $\langle\Psi|\Psi\rangle =1$ and the de-excitation $\Lambda$ operator can be determined from solving the left-eigenvalue problem for the similarity 
transfromed Hamiltonian $\overline{H}$
\begin{equation}
\langle\Phi|(1+\Lambda) \overline{H} (P^{(N)}+Q^{(N)}) = E_{\rm CC}^{(N)}\langle\Phi|(1+\Lambda)(P^{(N)}+Q^{(N)}) ~. 
\label{lambdaeq}
\end{equation}

In analogy to the ``right'' eigenvalue problem discussed in the previous Section, i.e. we can decompose  for a given de-excitation SES $\mathfrak{h}$ 
the exact $\Lambda$ operator from Eq. (\ref{lambdaeq}) in the same way as  
the $T$ operator (see Eq. (\ref{Tdecomp}) and drop off the explicit $\mathfrak{h}$-dependence for the simplification of notation), 
\begin{equation}
\Lambda = \Lambda_{\rm int}+\Lambda_{\rm ext}, ~~\text{with}~~[\Lambda_{\rm int},\Lambda_{\rm ext}]=0. \label{Ldecomp}
\end{equation}
In the exact wave function expansion, the left eigenvector of $\overline{H}$ can be alternatively represented in the extended CC form~\cite{arponen83_311}, 
\begin{equation}
\langle\Phi|(1+\Lambda) = \langle\Phi|e^S ~, \label{S}
\end{equation}
where the de-excitation cluster operator $S$ can also be decomposed  into commuting active and external parts 
\begin{equation}
S=S_{\rm int}+S_{\rm ext}, ~~\text{with}~~[S_{\rm int},S_{\rm ext}]=0. \label{sdecom}
\end{equation}
\textcolor{black}{In the exact wave function limit both parametrizations
are equivalent}
\begin{equation}
\langle\Phi|(1+\Lambda_{\rm int}+\Lambda_{\rm ext})=\langle\Phi|e^{S_{\rm int}} e^{S_{\rm ext}} ~.
\label{laes}
\end{equation}
After projecting onto $Q_{\rm int}^{(N)}$ and $Q_{\rm ext}^{(N)}$ subspaces one obtains the following identities
\begin{align}
\langle\Phi|(1+\Lambda_{\rm int})\mathcal{P}_{\rm int}^{(N)} &= \langle\Phi|e^{S_{\rm int}} \mathcal{P}_{\rm int}^{(N)} ~, \label{lsid1}  \\
\langle\Phi|\Lambda_{\rm ext}Q_{\rm ext}^{(N)} &= \langle\Phi|e^{S_{\rm int}} e^{S_{\rm ext}} Q_{\rm ext}^{(N)} ~. \label{lsid2} 
\end{align}
The cluster analysis of Eqs. (\ref{lsid1},\ref{lsid2}) gives
\begin{align}
\Lambda_{{\rm int},1} & = S_{{\rm int},1}  ~, \notag \\
\Lambda_{{\rm int},2} & = S_{{\rm int},2}+\frac{1}{2} S_{{\rm int},1}^2  ~, \notag \\
\Lambda_{{\rm int},3} & = S_{{\rm int},3}+ S_{{\rm int},1}S_{{\rm int},2}
+\frac{1}{6} S_{{\rm int},1}^3 ~, \label{ltos1} \\
 & \vdots   \notag 
\end{align}
and
\begin{align}
\Lambda_{{\rm ext},1} & = S_{{\rm ext},1}  ~, \notag \\
\Lambda_{{\rm ext},2} & = S_{{\rm ext},2}+\frac{1}{2} S_{{\rm ext},1}^2+S_{{\rm int},1}S_{{\rm ext},1}  ~, \notag \\
\Lambda_{{\rm ext},3} & = S_{{\rm ext},3}+ S_{{\rm ext},1}S_{{\rm ext},2}
+S_{{\rm int},1}S_{{\rm ext},2}\notag \\
&~~+S_{{\rm int},2}S_{{\rm ext},1}
+\frac{1}{2} S_{{\rm int},1}^2 S_{{\rm ext},1}  
 \notag \\
&~~ +\frac{1}{2} S_{{\rm int},1} S_{{\rm ext},1}^2+\frac{1}{6} S_{{\rm ext},1}^3 ~, \label{ltos2} \\
 & \vdots   \notag 
\end{align}
Reverse mapping can be readily derived from the above equations. 
Introducing exponential parametrizations (\ref{S}-\ref{lsid2}) into Schr\"odinger equation, one obtains
\begin{equation}
\langle\Phi|e^{S_{\rm int}} e^{S_{\rm ext}} \overline{H} e^{-S_{\rm ext}} = E_{\rm CC}^{(N)}\langle\Phi|e^{S_{\rm int}} ~. 
\label{bubu}
\end{equation}
Projecting these equations onto $\mathcal{P}_{\rm int}^{(N)}$ and $Q_{\rm ext}^{(N)}$ one gets the working equations for $e^{S_{\rm int}}$ and $S_{\rm ext}$ 
\begin{align}
\langle\Phi|e^{S_{\rm int}} \overline{\overline{H}}_{\rm ext}\mathcal{P}_{\rm int}^{(N)} &= E_{\rm CC}^{(N)}\langle\Phi|e^{S_{\rm int}}\mathcal{P}_{\rm int}^{(N)} ~, \label{lieq} \\
\langle\Phi|(1+\Lambda_{\rm int}) \overline{\overline{H}}_{\rm ext}Q_{\rm ext}^{(N)} &= 0 ~, \label{seeq}
\end{align}
where 
\begin{equation}
\overline{\overline{H}}_{\rm ext}=e^{S_{\rm ext}}\overline{H} e^{-S_{\rm ext}} = (e^{S_{\rm ext}}\overline{H})_C ~.
\label{doublebe}
\end{equation}
One can observe that, similar to Eq. (\ref{eff}), the energy $E_{\rm CC}^{(N)}$ and 
\begin{align}
    \langle \Psi_{\rm int} | = \langle \Phi | e^{S_{\rm int}}\mathcal{P}_{\rm int}^{(N)}
\end{align}
(and therefore $S_{\rm int}$ and $\Lambda_{\rm int}$) can be obtained by diagonalizing another effective many-body Hamiltonian 
\begin{align}
\overline{\overline{H}}_{\rm ext}^{\rm eff} = 
\mathcal{P}_{\rm int}^{(N)} \overline{\overline{H}}_{\rm ext} \mathcal{P}_{\rm int}^{(N)}~. \label{eff2}
\end{align}
Again, the amplitudes defining $\overline{\overline{H}}_{\rm ext}$  and corresponding to  $T$ and $S_{\rm ext}$ operators are decoupled from $S_{\rm int}$ or $\Lambda_{\rm int}$ amplitudes. 

\textcolor{black}{Note that from the above construction $\overline{\overline{H}}_{\rm ext}^{\rm eff}$ is generally not equal to $\overline{H}_{\rm ext}^{\rm eff}$, thus (i) $\langle \Psi_{\rm int}|\Psi_{\rm int} \rangle\neq 1$ and (ii) $\overline{\overline{H}}_{\rm ext}^{\rm eff}$ and $\overline{H}_{\rm ext}^{\rm eff}$ need to be constructed separately to solve both $T_{\rm int}$ and $\Lambda_{\rm int}$ (or $S_{\rm int}$). To address these issues, we can construct the third form of effective Hamiltonian based on the $\Lambda$-CC energy form \textcolor{black}{(in the exact wave function limit)}
\begin{align}
    E_{\rm CC}^{(N)} &= \langle \Phi | (1+\Lambda)\overline{H} | \Phi \rangle \notag \\
    &= \langle \Phi | e^{S_{\rm int}}e^{S_{\rm ext}}e^{-T_{\rm int}}\overline{H}_{\rm ext} e^{T_{\rm int}}| \Phi \rangle 
\end{align}
Inserting resolution of identity, $1=e^{-T_{\rm int}}e^{T_{\rm int}}$, between $e^{S_{\rm int}}$ and $e^{S_{\rm ext}}$, and define the effective Hamiltonian
\begin{align}
    \widetilde{H}^{\rm eff}_{\rm ext} &= \mathcal{P}_{\rm int}^{(N)} W \overline{H}_{\rm ext} \mathcal{P}_{\rm int}^{(N)},\\
    W &= e^{T_{\rm int}}e^{S_{\rm ext}}e^{-T_{\rm int}},
\end{align}
one gets another working equation for $S_{\rm int}$ and $T_{\rm int}$
\begin{align}
    E_{\rm CC}^{(N)} = \langle \Phi | e^{S_{\rm int}}e^{-T_{\rm int}} \widetilde{H}^{\rm eff}_{\rm ext} e^{T_{\rm int}} | \Phi \rangle,
\end{align}
where the left and right eigenvectors of $\widetilde{H}^{\rm eff}_{\rm ext}$ are given by
\begin{align}
    \langle \Psi_{\rm int} | = \langle \Phi | e^{S_{\rm int}}e^{-T_{\rm int}},~~ | \Psi_{\rm int} \rangle = e^{T_{\rm int}} | \Phi \rangle.
\end{align}
and $\langle \Psi_{\rm int} | \Psi_{\rm int} \rangle = 1$. The construction and use of the above three effective Hamiltonians are summarized in Tab. \ref{tab:Heff}
\begin{table*}[ht]
\begin{tabular}{lccc}
\hline \hline
 Effective Hamiltonian & $E_{\rm CC}^{(N)}$ & $T_{\rm int}$ & $S_{\rm int}$ \\ \hline
&&&\\
\multirow{2}{*}{$\overline{H}_{\rm ext}^{\rm eff} = \mathcal{P}_{\rm int}^{(N)} \overline{H}_{\rm ext} \mathcal{P}_{\rm int}^{(N)}$} &  
\multirow{2}{*}{$\surd$} &
\multirow{2}{*}{$|\Psi_{\rm int}\rangle = \mathcal{P}_{\rm int}^{(N)} e^{T_{\rm int}} | \Phi \rangle$} & 
\multirow{2}{*}{$\bigtimes$}  \\
&&&\\
\multirow{2}{*}{$\overline{\overline{H}}_{\rm ext}^{\rm eff} = \mathcal{P}_{\rm int}^{(N)} \overline{\overline{H}}_{\rm ext} \mathcal{P}_{\rm int}^{(N)}$} & 
\multirow{2}{*}{$\surd$} &
\multirow{2}{*}{$\bigtimes$} & \multirow{2}{*}{$\langle \Psi_{\rm int} | = \langle \Phi | e^{S_{\rm int}}\mathcal{P}_{\rm int}^{(N)}$}  \\
&&&\\
\multirow{2}{*}{$\widetilde{H}^{\rm eff}_{\rm ext} = \mathcal{P}_{\rm int}^{(N)} W \overline{H}_{\rm ext} \mathcal{P}_{\rm int}^{(N)}$} & 
\multirow{2}{*}{$\surd$} &
\multirow{2}{*}{$|\Psi_{\rm int}\rangle = \mathcal{P}_{\rm int}^{(N)} e^{T_{\rm int}} | \Phi \rangle$} & 
\multirow{2}{*}{$\langle \Psi_{\rm int} | = \langle \Phi | e^{S_{\rm int}} e^{-T_{\rm int}}\mathcal{P}_{\rm int}^{(N)}$}  \\
&&&\\
\hline 
\end{tabular}
\caption{The construction of three SES-CC effective Hamiltonians and their use to get $T_{\rm int}$ and $S_{\rm int}$.\label{tab:Heff}}
\end{table*}
}


\subsection{Green's function embedding based on the sub-system excitation/de-excitation sub-algebras} \label{SESCCGF}

In this section we describe a procedure for quantum description of the composite system based on the integration of quantum region (or embedded region) with the environment using Green's function formalism. Our strategy is valid for sub-systems defined by appropriate active spaces. In the spirit of quantum flow formulation \cite{kowalski2021dimensionality}, the SES-CC formalism assures a number of important properties which are ideally suited for  multi-center quantum embedding. 

Let's assume that all active  orbitals $\lbrace  \phi_M  \rbrace \in (\mathcal{R},\mathcal{S})$
defining $g^{(N)}(\mathcal{R},\mathcal{S})$ 
and $g^{(N)}(\mathcal{R},\mathcal{S})^{\dagger}$ that determine $T_{\rm int}$ and $\Lambda_{\rm int}$ operators are defined such that in 
the non-interacting limit (NLS) when composite system AB is separated into non-interacting sub-systems A and B all $g^{(N)}(\mathcal{R},\mathcal{S})$/$g^{(N)}(\mathcal{R},\mathcal{S})^{\dagger}$
orbitals evolve to orbitals localized  on the sub-subsystem A while the remaining orbitals evolve into orbitals localized on sub-system B. 

\textcolor{black}{In this section, we will analyze the asymptotic limit of the sub-block $G_{KL}(\omega)$ within the entire CC Green’s function matrix. This sub-block corresponds to spinorbitals $K$ and $L$, which define the $g^{(N)}(\mathcal{R},\mathcal{S})$/$g^{(N)}(\mathcal{R},\mathcal{S})^{\dagger}$ sub-algebras in an embedding region. In the NLS, these evolve into the $K_A$ and $L_A$ spinorbitals, localized on subsystem $A$,}
\begin{equation}
K  \xrightarrow{NSL} K_A  ~~, L \xrightarrow{NSL} L_A ~.
\label{paqa}
\end{equation}
Without loss of generality we will focus on the $\omega$-dependent CC Green's function $G_{KL}(\omega)$,
\begin{align}
&G_{KL}(\omega) = \notag \\
&\langle\Phi|(1+\Lambda) \overline{a_L^{\dagger}} X_K(\omega)|\Phi\rangle + \langle\Phi|(1+\Lambda)\overline{a_K} Y_L(\omega)|\Phi\rangle~,
\end{align}
where $\overline{a_K}$ and $\overline{a_L^\dagger}$ operators are defined as \cite{kowalski16_144101,kowalski16_062512}
\begin{align}
\overline{a_K} &= e^{-T} a_K e^T = a_K + [a_K,T], \notag \\  
\overline{a_L^\dagger} &= e^{-T} a_L^\dagger e^T = a_L^\dagger + [a_L^\dagger,T]~,
\end{align}
and $X_K(\omega)$ and $Y_L(\omega)$ satisfy the equations
\begin{align}
Q^{(N-1)} (\omega_{-i0}+\overline{H}_N) X_K(\omega) |\Phi\rangle &=
Q^{(N-1)} \overline{a_K} |\Phi\rangle ~, \label{x} \\
Q^{(N+1)} (\omega_{+i0}-\overline{H}_N) Y_L(\omega) |\Phi\rangle &=
Q^{(N+1)} \overline{a^\dagger_L} |\Phi\rangle ~, \label{y} 
\end{align}
First, using decomposition (\ref{laes},\ref{lsid1}), i.e., 
\begin{equation}
\langle\Phi|(1+\Lambda) = \langle\Phi|(1+\Lambda_{\rm int}) e^{S_{\rm ext}}~, \label{lblo}
\end{equation}
we can represent sub-block matrix elements $G_{KL}(\omega)$ as 
\begin{align}
G_{KL}(\omega) &= \langle\Phi|(1+\Lambda_{\rm int}) e^{S_{\rm ext}} \overline{a_L^{\dagger}} X_K(\omega)|\Phi\rangle \notag \\
&+ \langle\Phi|(1+\Lambda_{\rm int}) e^{S_{\rm ext}}\overline{a_K} Y_L(\omega)|\Phi\rangle ~,
\label{GKL}
\end{align}
Next, if we formally decompose the $X_K(\omega)$ and $Y_L(\omega)$ operators into their internal and external parts  
\begin{align}
X_K(\omega)&=X_{K,{\rm int}}(\omega)+X_{K,{\rm ext}}(\omega)~, \\
Y_L(\omega)&=Y_{L,{\rm int}}(\omega)+Y_{L,{\rm ext}}(\omega)~,
\end{align}
with the internal parts, $X_{K,{\rm int}}(\omega)$ and $Y_{L,{\rm int}}(\omega)$, reconstruct the correlation effects in the embedded  sub-system (in $(N-1)$- and $(N+1)$-electron Hilbert spaces, respectively) described by  active orbitals 
whereas the external parts, $X_{K,{\rm ext}}(\omega)$ and $Y_{L,{\rm ext}}(\omega)$, describe correlation effects in the environment and effects that simultaneously  correlate embedded system and the environment, we can then decompose the $G_{KL}(\omega)$ as follows
\begin{widetext}
\begin{align}
G_{KL}(\omega)&=G^{\rm int}_{KL}(\omega)+G^{\rm ext}_{KL}(\omega) ~, \label{gdec} \\
G^{\rm int}_{KL}(\omega) &= \langle\Phi|(1+\Lambda_{\rm int})\overline{a_L^{\dagger}} X_{K,{\rm int}}|\Phi\rangle + \langle\Phi|(1+\Lambda_{\rm int})\overline{a_K} Y_{L,{\rm int}}|\Phi\rangle ~,
\label{gsemb} \\
G^{\rm ext}_{KL}(\omega) &= 
\langle\Phi|(1+\Lambda_{\rm int})(e^{S_{\rm ext}}-1)\overline{a_L^{\dagger}} 
(X_{K,{\rm int}}+X_{K,{\rm ext}})|\Phi\rangle + \langle\Phi|(1+\Lambda_{\rm int})(e^{S_{\rm ext}}-1)\overline{a_K} 
(Y_{L,{\rm int}}+Y_{L,{\rm ext}})|\Phi\rangle ~.
\label{gsint}
\end{align}
\end{widetext}

As shown in Appendices A, B, C which analyze the asymptotic behavior of $T$, $\Lambda_{\rm int}$, $S_{\rm ext}$, $X_K(\omega)$, and $Y_K(\omega)$ operators, one can demonstrate (see Appendix C) that $G^{\rm int}_{KL}(\omega)$ evolves in the NSL to the CC Green's function for isolated subsystem A. Therefore, we identify the $G^{\rm int}_{KL}(\omega)$ part describing embedded subsystem A while the $G^{\rm ext}_{KL}(\omega)$ part describes the interactions of the embedded system with the surrounding environment.

Our embedding procedure is based on the construction of the following representation of the Green's function, 
\begin{align}
    \mathbf{G}(\omega) = 
    \left[\begin{array}{c|c} \mathbf{G}^{\rm emb}(\omega) & \mathbf{G}^{\rm env,1}(\omega)\\ \hline
    \mathbf{G}^{\rm env,2}(\omega) & \mathbf{G}^{\rm env,3}(\omega) \end{array} \right], \label{Gblk1}
\end{align}
where the embedded block $\mathbf{G}^{\rm emb}(\omega)$ and the environment blocks ($\mathbf{G}^{\rm env,1}(\omega)$, $\mathbf{G}^{\rm env,2}(\omega) $, and $\mathbf{G}^{\rm env,3}(\omega)$)
are defined by matrix elements $G_{KL}(\omega)$,  $G_{K\overline{p}}(\omega)$, $G_{\overline{p}K}(\omega)$, and $G_{\overline{p}\overline{q}}(\omega)$, respectively. Here 
$\overline{p}$ and $\overline{q}$ designate inactive spinorbitals. 
 
For several active-space problems such as the quantum flow equations discussed for CC formalism in Ref.\citenum{kowalski2021dimensionality}, schematic representation of the Green's function featuring multiple embedded parts ($\mathbf{G}^{\rm emb(n) }(\omega)$, $n=1,\cdots,M$, $M\ge2$ ) is shown below:
\begin{align}
    \mathbf{G}(\omega) = 
    \left[\begin{array}{c|c|c} \mathbf{G}^{\rm emb(1) }(\omega) & \mathbf{G}^{\rm env}(\omega) & \cdots \\ \hline
    \mathbf{G}^{\rm env }(\omega) & \mathbf{G}^{\rm emb(2)}(\omega) & \cdots \\ \hline
    \vdots & \vdots & \ddots \end{array}  \right],\label{Gblk2}
\end{align} 

\textcolor{black}{Note that in Eq. (\ref{Gblk2}), the Green's function of the full system, including both the embedding and the environment subblocks, is treated on an equal footing. This approach essentially involves a re-partitioning of the CC Green's function for the full system, based on the partitioning of the excitation manifold. For systems described by the CC method, the Hamiltonian commutes with the particle number operator. As the CC method is an approximation to solve the Schrödinger equation with this Hamiltonian, the approximation maintains this fundamental commutation relationship, thereby conserving particle number. It is also worth noting that the total energy can be computed both from Green's function methods and from wave function theories. In the full configuration interaction limit, the total energies computed from both approaches should theoretically converge to the same value, assuming that all relevant correlations and interactions are fully accounted for. However, in approximate approaches, the total energy computed from both methods generally is not equal. A recent discussion from the perspective of the Luttinger-Ward $\Phi$-functional can be found in Ref. \citenum{shee2019CC}. Briefly, in $\Phi$-derivable approaches~\cite{PhysRev.118.1417}, the self-energy is derived as a functional derivative of a $\Phi$-functional with respect to the Green's function. This approach ensures certain conservation laws (related to particle number, energy, momentum, etc.) are inherently satisfied. Consequently, quantities like electronic energy can be evaluated in multiple ways (e.g., through thermodynamic integration or the Galitskii-Migdal formula~\cite{galitskii1958application}) and yield the same result. On the other hand, CC theory is not formulated as a diagrammatic expansion of a $\Phi$-functional. Therefore, the CC Green's function, and consequently the CC self-energy, is not a functional derivative of a $\Phi$-functional. This means that CC theory does not inherently guarantee the same kind of diagrammatic consistency as $\Phi$-derivable methods. As a result, different approaches to calculating quantities like electronic energy may not necessarily yield identical results.}

The properties discussed in Appendices \ref{appendix1}, \ref{appendix2}, \ref{appendix3} guarantee that in NSL the above matrix assumes block-form with blocks defined for embedding regions and the environment, where $X_{K{\rm(i), int(i)}}(\omega)$ and $X_{K{\rm(i), ext(i)}}(\omega)$, as well as $Y_{L{\rm(i), int(i)}}(\omega)$ and $Y_{L{\rm(i), ext(i)}}(\omega)$, are defined through
\begin{widetext}
\begin{align}
Q_{\rm int(i)}^{(N-1)} [\omega_{-i0}+\overline{H}_N]  (X_{K{\rm(i), int(i)}}(\omega)+X_{K{\rm(i), ext(i)}}(\omega)) |\Phi\rangle &= 
Q_{\rm int(i)}^{(N-1)} \overline{a_{K(i)}} |\Phi\rangle ~, (i=1,\ldots,M) \label{aaxint} \\
Q_{\rm int(i)}^{(N+1)} [\omega_{+i0}-\overline{H}_N]  (Y_{L{\rm(i), int(i)}}(\omega)+Y_{L{\rm(i), ext(i)}}(\omega)) |\Phi\rangle &=
Q_{\rm int(i)}^{(N+1)} \overline{a^{\dagger}_{L(i)}} |\Phi\rangle ~, (i=1,\ldots,M)\label{aayint}
\end{align}
\end{widetext}
where the index $i$ goes over all active centers (active spaces). 
The flow-type representation of the CC Green's function embedding in (\ref{Gblk1}) and (\ref{Gblk2}) offers a flexible selection mechanism for cluster amplitudes included 
in the $N$-electron problem. As discussed in Ref.  \citenum{kowalski2021dimensionality} the cluster operator $T$ (the $\Lambda$ operator is defined in a similar way) is defined as a combination of unique
cluster amplitudes defining the set of internal cluster operators for each active space involved in the flow. The choice of the active spaces may be dictated by the nature of processes of interest. For example, if one is interested in describing transport processes, the active spaces should be chosen to cover the spatial area relevant to the junction. For other applications, such as calculating core-level binding energies, the active spaces in the flow should be chosen to provide a way of correlating core orbitals with valence orbitals, which is also important in bringing a balance between ground-state and core-level states. The flow-type representation of the CC Green's function also offers a possibility of representing the entire problem in the form of small-dimensionality 
computational blocks. 
Remarkably, for various active space SES-CC problem either $\overline{H}_{\rm ext}^{\rm eff}$ or $\overline{\overline{H}}_{\rm ext}^{\rm eff}$ is providing form of the effective interactions that can be used to calculate the exact energy of the $N$-electron system (as long as $T_{\rm ext}$ is known) in the active space. 
In this context, take $\overline{H}_{\rm ext}^{\rm eff}$ for an example, by diagonalizing this effective Hamiltonian, we get $E_{\rm CC}^{(N)}$ and its bi-orthonormal left and right eigenvectors, $\langle \Psi_{\rm int}|$ and $|\Psi_{\rm int} \rangle$, that bypass the $\Lambda$-parameterization but satisfy $\langle \Psi_{\rm int}|\Psi_{\rm int} \rangle=1$. We could then reformulate the sub-block (\ref{GKL}) as
\begin{widetext}
\begin{align}    
    G_{KL} &=     
    \langle \Psi_{\rm int}|(P^{(N)}+ Q_{\rm int}^{(N)})W\overline{a_{L}^{\dagger}}_{,\rm ext} Q_{\rm int}^{(N-1)} (\omega_{-i0^+} +\overline{H}_{N,\rm ext} )^{-1} Q_{\rm int}^{(N-1)} \overline{a_{K}}_{,\rm ext} (P^{(N)}+ Q_{\rm int}^{(N)})| \Psi_{\rm int} \rangle    \notag \\
    &+ \langle \Psi_{\rm int}| (P^{(N)}+ Q_{\rm int}^{(N)})W\overline{a_{K}}_{,\rm ext} Q_{\rm int}^{(N+1)}(\omega_{+i0^+} -\overline{H}_{N,\rm ext} )^{-1} Q_{\rm int}^{(N+1)} \overline{a_{L}^{\dagger}}_{,\rm ext} (P^{(N)}+ Q_{\rm int}^{(N)})| \Psi_{\rm int} \rangle, \label{GKL2}
\end{align}
\end{widetext}
where $\overline{H}_{N,\rm ext} = \overline{H}_{\rm ext} - \langle \overline{H}_{\rm ext} \rangle $
(in this paper we assume for certain systems, the $\langle \overline{H}_{\rm ext} \rangle$ can be approximated by $E_{\rm CC}^{(N)}$), and 
\begin{align}
    \overline{a_{K}}_{,\rm ext} &= e^{-T_{\rm ext}} a_K e^{T_{\rm ext}}~, 
    ~~\overline{a_{L}^{\dagger}}_{,\rm ext} = e^{-T_{\rm ext}} a_L^\dagger e^{T_{\rm ext}} ~.
\end{align}
Note that (i) in (\ref{GKL2}) $\langle \Psi_{\rm int}|$ essentially approximates $\langle \Phi | (1+\Lambda)e^{-T_{\rm int}}$, such that the projection $(P^{(N)}+Q_{\rm int}^{(N)})$ can be applied between $\langle \Psi_{\rm int}|$ and $\overline{a_{K}}_{,\rm ext}$ (or $\overline{a_{L}^{\dagger}}_{,\rm ext}$), (ii) the pole positions associated with the effectiven Hamiltonian are exactly part of the spectrum of similarity-transformed Hamiltonian that is associated with the active space, and (iii) the similar reformulation can also be obtained for the unitary version of CC Ansatz. Note that the unitary CC effective Hamiltonians are expressed in terms of anti-Hermitian cluster operators which involve non-terminating expansions in exact limit, and thus require approximate forms in practice. In the context of double unitary CC (DUCC) based effective Hamiltonian \cite{Bauman2019SESDUCC,kowalski2020sub,kowalski2021dimensionality,bauman2021coupled}, approximate effective Hamiltonian forms involving single- and double-commutator expansions have been calibrated for reproducing exact ground state energy using small active spaces, and thus showing potential for working with Green's function embedding approaches. Relevant discussions of this topic will be presented in our future work.


\section{Numerical results and discussion}\label{sec:numerical}

\subsection{SES-CCGF workflow}

In practical applications approximate schemes of SES-CCGF formalism can be driven by the truncation of excitation levels used to define these operators or by using low-order perturbative approximations. In this section, we demonstrate the SES-CCGF workflow by computing the spectral function of a single-impurity Anderson model (SIAM) with three electrons and two bath site, $N_{\rm bath}=2$. The SIAM is a well-established model that describes a system consisting of a localized impurity with a repulsive interaction between opposite spin electrons coupled with a non-interacting electronic bath. \cite{Anderson1961SIAM, Newns1969SIAM, Wilson1075RMP} This model has been successfully employed to explain the metal-insulator transition through Anderson localization. Here, we apply SES-CCGF with single and double excitations to the SIAM to demonstrate how SES-CCSDGF works and how to apply further approximation.

\begin{figure}
    \centering
    \includegraphics[width=\linewidth]{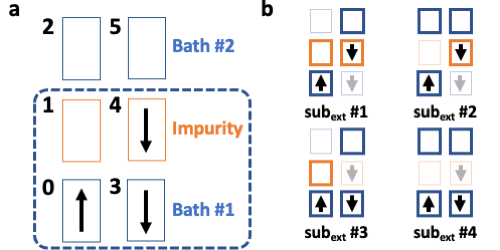}
    \caption{Electronic structure of the three-site SIAM employed in the present work. The embedded/active regime including the impurity site and one bath site is framed. Numbers $0-5$ label the six spin sites. The electron configuration with one up-spin electron on site `0' and two down-spin electrons on sites `3' and `4' is employed as the reference state.}
    \label{3siteAIM}
\end{figure}

The SIAM Hamiltonian comprises three parts
\begin{align}
H_{\rm SIAM} = H_{\rm imp.} + H_{\rm bath} + H_{hyb.}, \label{Hsiam}
\end{align}
where 
\begin{align}
H_{\rm imp.} = \sum_{\sigma} (\epsilon_c - \mu) c^\dagger_{\sigma} c_{\sigma} + U c^\dagger_{\uparrow}c_{\uparrow}c^\dagger_{\downarrow}c_{\downarrow} \label{Himp}
\end{align}
describes the impurity-site energy $\epsilon_c$ and the Coulomb interaction $U$ between the electrons with opposite spins ($\sigma = \uparrow$ or $\downarrow$) at the impurity site,
\begin{align}
H_{\rm bath} = \sum_{i=1,\sigma}^{N_{\rm b}} \epsilon_{d,i} d^\dagger_{i,\sigma} d_{i,\sigma} \label{Hbath}
\end{align}
describes the non-interacting bath site with $\epsilon_{d}$'s being the bath-site energies, and
\begin{align}
H_{\rm hyb.} = \sum_{i=1,\sigma}^{N_{\rm b}} V_i \big( c^\dagger_{\sigma} d_{i,\sigma} + d^\dagger_{i,\sigma} c_{\sigma} \big) \label{Hhyb}
\end{align}
describes the coupling between the impurity site and the bath levels due to the hybridization. 
The electronic structure of the three-site SIAM is exhibited in Fig. \ref{3siteAIM}, where we select one impurity site and one bath site as embedded region. The reference state is $|\Phi\rangle = |100110\rangle$ where we include six spin sites with `1' denoting occupied and `0' denoting virtual status of the corresponding spin site. 

\begin{table*}
\begin{tabular}{c|rrrrrrrr}
\hline \hline 
Determinants& $|\Phi_0^1\rangle$       & $|\Phi_0^2\rangle$       & $|\Phi_3^5\rangle$       & $|\Phi_4^5\rangle$       
            & $|\Phi_{03}^{15}\rangle$ & $|\Phi_{04}^{15}\rangle$ & $|\Phi_{03}^{25}\rangle$ & $|\Phi_{04}^{25}\rangle$ \\
           \hline
$t$-amplitudes 
            & -0.628627 & 0.244428 & -0.244428 & -0.628627 
            &  0.013884 & 0.093685 & -0.003991 & -0.013884  \\ 
$\lambda$-amplitudes 
            & -0.442227 & 0.165380 & -0.165380 & -0.442227 
            &  0.075843 & 0.221301 & -0.028853 & -0.075843  \\
$s$-amplitudes 
            & -0.442227 & 0.165380 & -0.165380 & -0.442227 
            &  0.002707 & 0.025736 & -0.001502 & -0.002707  \\
                     \hline \hline 
\end{tabular} 
\caption{CCSD amplitudes ($t$'s, $\lambda$'s, and $s$'s) of three-site SIAM model Hamiltonian with $\epsilon_c = -0.5$, $\epsilon_d = \{-1.0,1.0\}$, $U = 1.0$, and $V_i=\{1.0, 1.0\}$.\label{tab:CC_TAmp}}
\end{table*}

\begin{figure*}
    \centering
    \includegraphics[width=\linewidth]{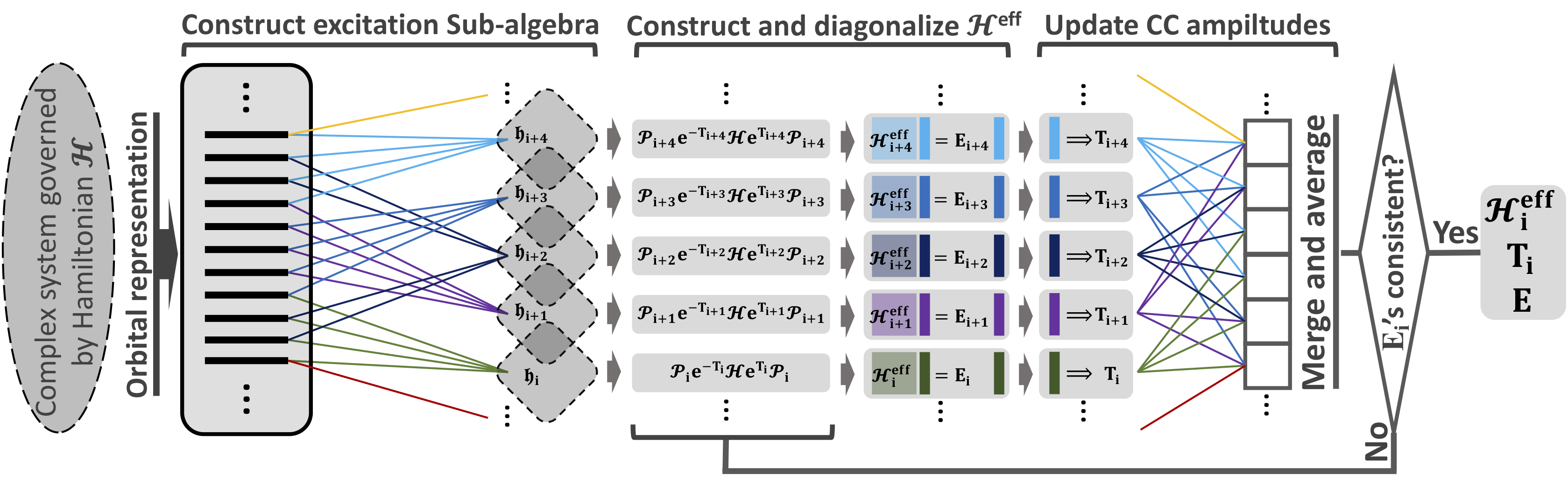}
    \caption{SES-CC workflow to obtain the ground state energy, effective hamiltonian $\overline{H}^{\rm eff}$ and cluster amplitude corresponding to the subsystems. When converge, the energy of each subsystem  matches the energy of the entire system.
    \label{fig:flow}}
\end{figure*}

\subsection{Ground state energy}

%
%
Given a three-site SIAM Hamiltonian with $U = 1.0$, $\epsilon_c = -0.5$, $\epsilon_d = \{-1.0, 1.0\}$, and $V_i=\{1.0,1.0\}$, $E_{\rm 0}=-3.7572543$ a.u. Note that since only single and double excitations could occur in the model system, the CCSD calculation reproduces the true ground state energy with the computed CC $T$ amplitudes given by
\begin{align}
    T &= T_{\rm int} + T_{\rm ext} \notag \\
    T_{\rm int} &= t_0^1 a_1^\dagger a_0 \notag \\
    T_{\rm ext} &= t_0^2 a_2^\dagger a_0 + t_3^5 a_5^\dagger a_3 + t_4^5 a_5^\dagger a_4 + t_{03}^{15} a_1^\dagger a_5^\dagger a_3 a_0 \notag \\ 
    &~~+ t_{04}^{15} a_1^\dagger a_5^\dagger a_4 a_0 + t_{03}^{25} a_2^\dagger a_5^\dagger a_3 a_0 + t_{04}^{25} a_2^\dagger a_5^\dagger a_4 a_0 .
\end{align}
All the CCSD amplitudes ($t$'s, $\lambda$'s, and $s$'s) are given in Tab. \ref{tab:CC_TAmp}

\begin{figure*}
    \centering
    \includegraphics[width=\linewidth]{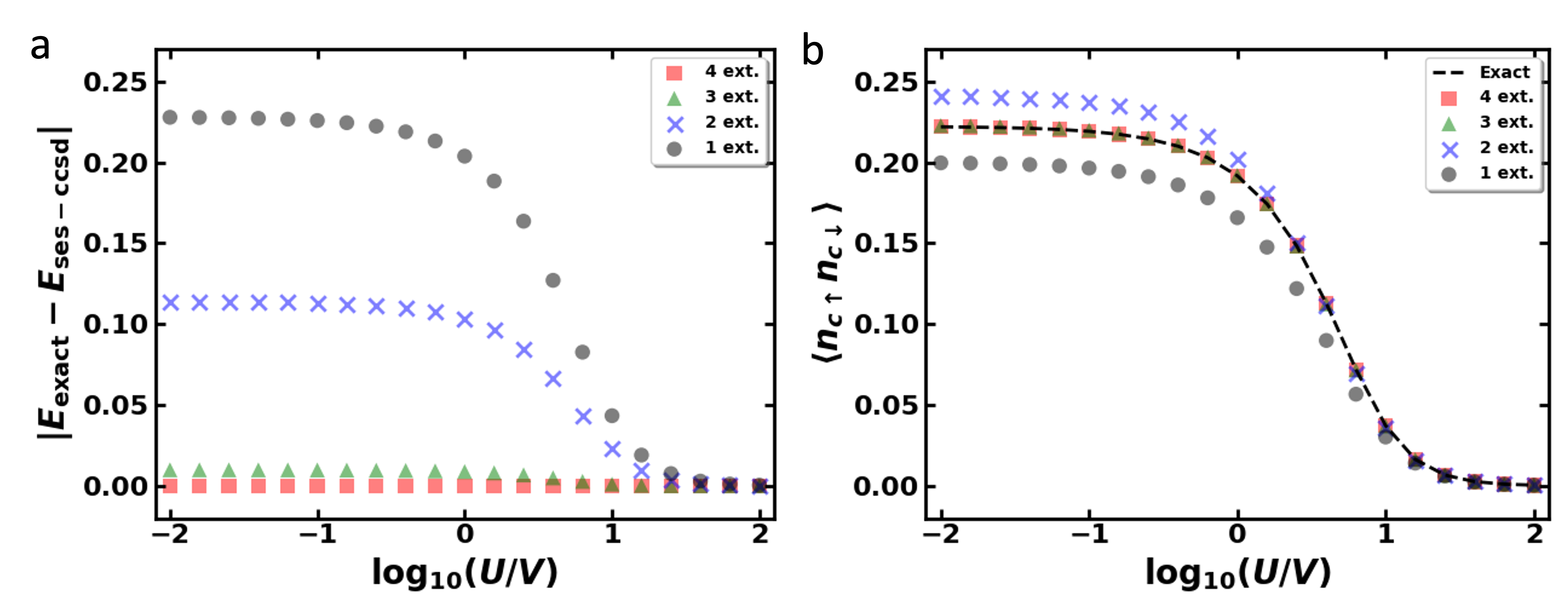}
    \caption{(a) Energy deviation of SES-CCSD ground-state energy with respect to exact ground-state energy SIAM and (b) double occupancy of the impurity $\langle n_{c\uparrow}n_{c\downarrow}\rangle$ computed by SES-CCSD as functions of $U$ for the symmetric SIAM with different numbers of external subsystems. In the three-site SIAM Hamiltonian, we select $V=1$, $\epsilon_c = -U/2$, $\epsilon_d=\{-1.0,1.0\}$.}
    \label{fig:En_occ_3site}
\end{figure*}

\subsection{SES-CC effective Hamiltonian}

By employing $T_{\text{ext}}$, we can downfold the $2^6 \times 2^6$ three-site SIAM Hamiltonian to the $2 \times 2$ effective Hamiltonian, $\overline{H}^{\rm eff}_{\rm ext}$, using Eq. (\ref{eff}). In practical calculations, we propose an iterative computational procedure for performing the SES-CC embedding calculations, as demonstrated in Fig. \ref{fig:flow}. This approach is similar to the recently proposed flow strategy~\cite{kowalski2021dimensionality,kowalski2023quantum} but operates within the conventional nonunitary coupled cluster framework. As depicted in Fig. \ref{fig:flow}, the proposed procedure comprises three basic steps. Initially, the configuration space of the system of interest is partitioned (potentially with overlaps) into subsystems based on various SESs. Subsequently, the projected effective Hamiltonians associated with the SESs are constructed and diagonalized. Finally, the eigenvector of each projected effective Hamiltonian is mapped back to the corresponding global positions associated with the coupled cluster amplitude tensor of the full system, followed by an amplitude synchronization and update. It is important to note that the second and third steps form a loop to iterate and update the coupled cluster amplitude. Upon convergence, the eigenvalues of each projected effective Hamiltonian become consistent and match the coupled cluster energy of the full system. For current cast, upon converging, we obtained the $2\times 2$ effective Hamiltonian as follows
\begin{align}
    \overline{H}_{\rm ext}^{\rm eff} &= (P^{(N)}+Q_{\rm int}^{(N)}) \overline{H}_{\rm ext} (P^{(N)}+Q_{\rm int}^{(N)}) \notag \\
    &= \left(\begin{array}{cc}
        -3.12862715 &  1.0 \\
         1.33811282 & -1.62862715
    \end{array}\right) \notag \\
    &= -2.37862715 I + 1.16905641 \sigma_x \notag \\
    &~~+ 0.16905641j \sigma_y -0.75 \sigma_z
\end{align}
where I is identity matrix, $\sigma_{x,y,z}$ are spin-1/2 Pauli matrices, and we define
\begin{align}
    Q_{\rm int}^{(N)} &= |\Phi_0^1\rangle\langle\Phi_0^1| , \notag \\
    Q_{\rm ext}^{(N)} &= |\Phi_0^2\rangle\langle\Phi_0^2| + |\Phi_3^5\rangle\langle\Phi_3^5| + |\Phi_4^5\rangle\langle\Phi_4^5| + |\Phi_{03}^{15}\rangle\langle\Phi_{03}^{15}|  \notag \\
    &~~ + |\Phi_{04}^{15}\rangle\langle\Phi_{04}^{15}| + |\Phi_{03}^{25}\rangle\langle\Phi_{03}^{25}| + |\Phi_{04}^{25}\rangle\langle \Phi_{04}^{25}|,
\end{align}
with the superscripts (subscripts) in the bra/ket labelling virtual (occupied) orbitals as showed in Fig. \ref{3siteAIM}. Diagonalizing $\overline{H}_{\rm ext}^{\rm eff}$ gives rise to the lowest eigenvalue $E_{\rm CC}^{(N)}$ (i.e. $E_0$) and the corresponding left and right eigenvectors, i.e $\langle \Psi_{\rm int}|$ and $|\Psi_{\rm int}\rangle$
\begin{align}
    |\Psi_{\rm int}\rangle &= e^{T_{\rm int}} |\Phi\rangle = |\Phi\rangle + t_0^1 |\Phi_0^1\rangle 
\end{align}
whose normalized form after projecting onto the active space gives
\begin{align}
    \left(0.8497802, -0.53419491 \right)^{\rm T}. \notag
\end{align}
Similarly, we have $\langle \Psi_{\rm int} |$ whose normalized form after projecting onto the active space gives
\begin{align}
    \left(0.90848166, -0.42679229\right). \notag
\end{align}
It's worth mentioning that, based on the proposed workflow, different levels of approximation can be introduced through the incrementation of the external subsystem, which would be particularly useful when the external degrees of freedom is too larger to deal with while the only limited number of them are pivotal for the accurate interaction between the internal system of interest and the external environment. For instance, simple demonstrations are shown in Figs. \ref{fig:En_occ_3site} for the three-site SIAM, where four external subsystems are included in an incremental manner to offer varying levels of approximation for energy and double occupation. 

\subsection{Spectral function}

\begin{figure*}
    \centering
    \includegraphics[width=\linewidth]{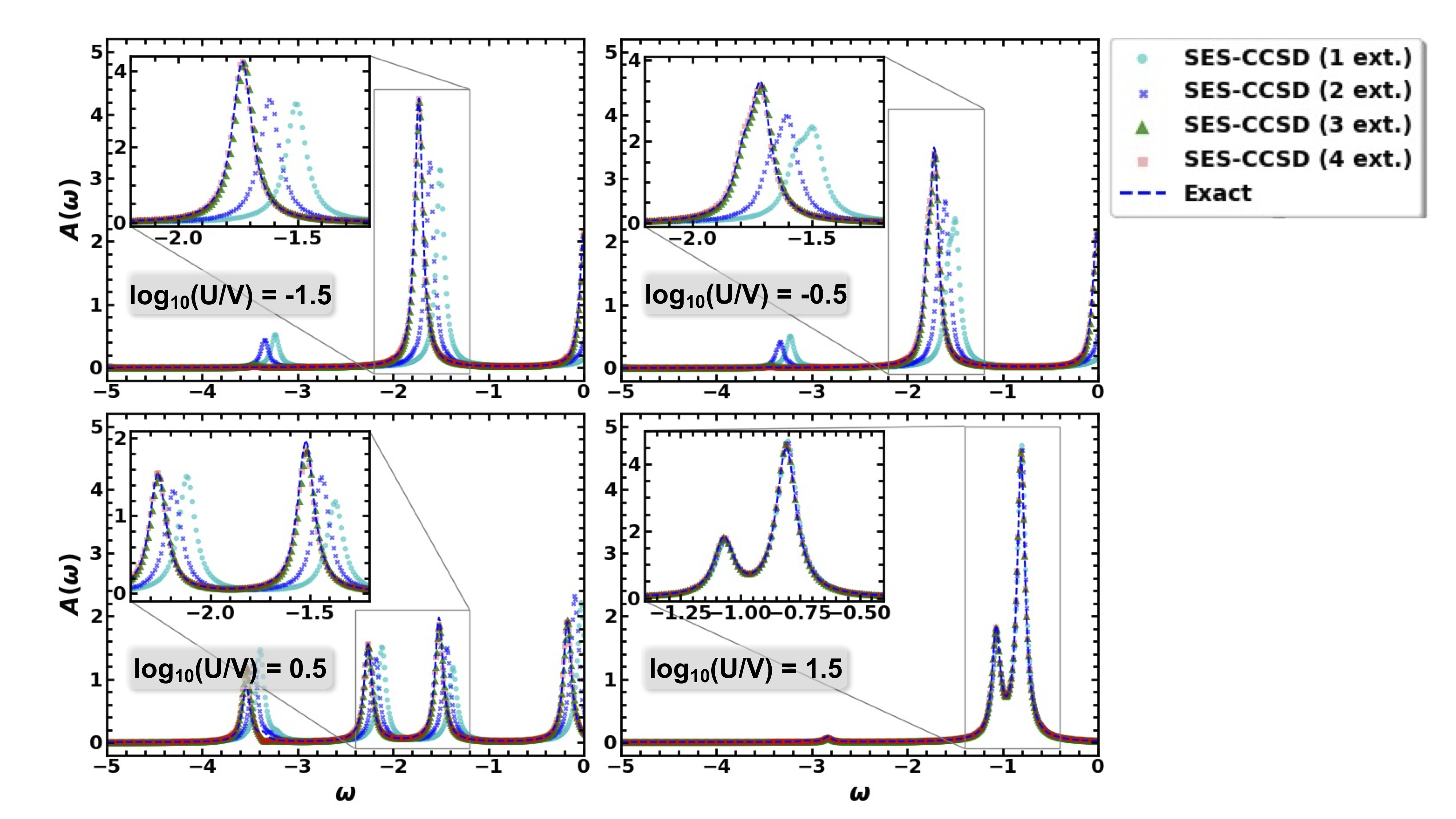}
    \caption{$\omega$-dependent spin-polarized spectral function, $A(\omega)$, of the three-site SIAM computed from exact diagonalization and CCSD approaches. In the three-site SIAM Hamiltonian, we select $V=1$, $\epsilon_c = -U/2$, $\epsilon_d=\{-1.0,1.0\}$.}
    \label{3site_full}
\end{figure*}

\textcolor{black}{The spectral function $A(\omega)$ can be computed from the Green's function through
\begin{align}
    A(\omega) = - \frac{1}{\pi} \text{Tr}[\Im \mathbf{G}(\omega)]
\end{align}
where the impurity Green's function matrix $\mathbf{G}(\omega)$ at the SES-CCSD level is computed employing Eq. (\ref{GKL2}) with 
\begin{align}
    Q_{\rm int}^{(N-1)} &= |\Phi_0^1\rangle\langle\Phi_0^1| + |\Phi_3^4\rangle\langle\Phi_3^4| , \notag \\  
    Q_{\rm int}^{(N+1)} &= \emptyset, 
\end{align}
and the effective Hamiltonians constructed using Eq. (\ref{eff}) over different choices of external subsystems. For comparison, we also computed $A(\omega)$ and impurity Green's function from the full spectrum obtained from the exact diagonalization of the SIAM Hamiltonian (\ref{Hsiam}).}
Fig. \ref{3site_full} shows the computed the impurity spectral function from exact diagonalization and SES-CCSDGF approach with different $U/V$ ratios ranging from weakly to strongly correlated scenarios. For the three-site model, the SES-CCSDGF workflow need to include all four external subsystems as listed in Fig. \ref{3siteAIM}b in order to  reproduces the exact spectra. 

\textcolor{black}{Regarding the computational cost, the SES-CC framework has a computational cost that is almost the same as a full CC calculation. However, it yields not only the same ground state energy but also an effective Hamiltonian, the dimension of which depends solely on the size of the active space (i.e., the number of excitations involved). The subsequent Green's function calculation is based on this dimension-reduced effective Hamiltonian, significantly reducing the computational cost compared to the full CCGF. For a general system, if the active space includes a constant number of orbitals, the computational cost of the Green's function calculation in SES-CCGF remains constant. However, the computational cost of generating the effective Hamiltonian may vary.}



\section{Conclusion and outlook}

In this paper, we have established a theoretical foundation for a novel and rigorous sub-system embedding within coupled cluster Green's function theory. Distinct from previously reported Green's function embedding methods~\cite{sun2016quantum}, our proposed theory relies on partitioning the excitation manifold through sub-system embedding subalgebras (SES). It operates directly on the $\omega$-dependent Green's function formalism within the conventional non-unitary coupled cluster theoretical framework. Our approach anticipates offering a meticulous description of system-environment coupling by utilizing existing coupled cluster machinery. Due to the non-unitary nature of conventional coupled cluster theory, an essential prerequisite for the proposed theory—which we have developed in this work—is the generalization of the SES-CC formalism to address the left eigenvalue problem through a specific form of Hamiltonian similarity transformation. This embedding approach is inherently suitable for multiscale/multiphysics simulations, enabling the iterative construction of multiple embedded sub-systems to balance accuracy and efficiency optimally.

Future studies could see the proposed theory directly linked to advancements in quantum computing. Specifically, an effective Hamiltonian—significantly reduced in dimension compared to the original—could be employed and more readily encoded within quantum computing frameworks for quantum dynamics simulations, or for isolating segments of the exact spectrum, assuming there are ample external subsystems to replicate the desired physics. Concerning the Green's function calculations within the proposed theoretical framework, the necessity arises to evaluate the inverse of the shifted projected Hamiltonian within the active space. To facilitate larger scale quantum simulations, further refinements may be derived from the recently suggested connected moment expansions (see Ref. \citenum{Peng2022State} for a recent review and the references therein), which have shown superior convergence performance compared to direct Taylor or geometric expansions. The primary motivation for this simplification is to employ straightforward and less complex circuits for expectation measurement, optimizing the number of measurements required through advanced techniques such as grouping or shadow techniques.

\section{Acknowledgements}

B.~P. and K.~K. were supported by  the ``Embedding QC into Many-body Frameworks for Strongly Correlated  Molecular and Materials Systems'' project, which is funded by the U.~S. Department of Energy, Office of Science, Office of Basic Energy Sciences (BES), the Division of Chemical Sciences, Geosciences, and Biosciences \textcolor{black}{under FWP 72689}. 

\section{Data Availability Statement}
The data that support the findings of this study are available from the corresponding author upon reasonable request.

\appendix

\section{Properties of $\Lambda_{\rm int}$ and  $S_{\rm ext}$ operators in the NSL} \label{appendix1}
Let us assume  that in the non-interacting sub-systems limit (NSL),  the entire composite system AB dissociates into sub-system A (defined by the active space generated by SES $\mathfrak{h}$) and the remaining part referred to as the sub-system B such that
\begin{align}
&H \xrightarrow{NSL}  H_A + H_B ~,~ |\Phi\rangle \xrightarrow{NSL}  |\Phi_A \Phi_B\rangle~, \notag \\
&E_{\rm CC}^{(N)} \xrightarrow{NSL} E_{A,\rm CC} + E_{B,\rm CC} ~,
\end{align}
which leads to $\overline{H}_{N} \xrightarrow{NSL}  \overline{H}_{N,A} + \overline{H}_{N,B}$.
In the further analysis, due to  of  definition of the internal and external parts of the operators, 
we will  also assume that 
\begin{align}
&T \xrightarrow{NSL} T_A + T_B  ~,
T_{\rm int}  \xrightarrow{NSL}  T_A ~, 
T_{\rm ext} \xrightarrow{NSL}  T_B ~, \notag \\
&\Lambda_{\rm int} \xrightarrow{NSL}  \Lambda_A ~.
\label{TNSL}
\end{align}
As discussed in Section \ref{LambdaSES}, the equations for $\Lambda_{\rm int}$ and $S_{\rm ext}$, in the exact wave function limit, can be cast in the form 
\begin{align}
\langle\Phi|(1+\Lambda_{\rm int}) [e^{S_{\rm ext}} \overline{H}_N e^{-S_{\rm ext}}] (P^{(N)}+Q_{\rm int}^{(N)}+Q_{\rm ext}^{(N)}) = 0 ~.
\label{seeq1}
\end{align}
Given the fact that $S_{\rm ext}$ cluster amplitudes  have to involve by  construction at least one spinorbital index on sub-system B, the general 
decomposition of $S_{\rm ext}$ in the NSL takes the form
\begin{equation}
S_{\rm ext} \xrightarrow{NSL} S_B + S_{AB}
\label{sabdec}
\end{equation} 
(where $S_{AB}$ is defined by the amplitudes indexed by both active and inactive spin-orbitals),
Eq. (\ref{seeq1}) can then be cast the form: 
\begin{widetext}
\begin{align}
\langle\Phi|(1+\Lambda_A) \left[e^{S_B+S_{AB}} \left(\overline{H}_{N,A}+\overline{H}_{N,B}\right) e^{-S_B-S_{AB}}\right] (P^{(N)}+Q_A^{(N)}+Q_B^{(N)}+Q_{AB}^{(N)}) = 0 ~,
\label{seeq2}
\end{align}
\end{widetext}
which can be re-written in the form 
\begin{align}
&\langle\Phi|(1+\Lambda_A) \left[\left(e^{S_{AB}} \overline{H}_{N,A}\right)_c+\left(e^{S_{AB}} \left(e^{S_B}\overline{H}_{N,B}\right)_c\right)_c\right] \notag \\
&(P^{(N)}+Q_A^{(N)}+Q_B^{(N)}+Q_{AB}^{(N)}) = 0 ~,
\label{seeq2}
\end{align}
where the subscript `$c$' denotes the connected parts of a given operator expression. Since the CC equations are satisfied for the subsystem, i.e., $Q_B\overline{H}_{N,B}|\Phi\rangle$ and all uncontracted lines emanating from $S_{AB}$ or $\overline{H}_{N,B}$ lead to vanishing expessions when acting onto reference function $|\Phi\rangle = |\Phi_A\Phi_B\rangle$ and excited configurations defining $Q_A$,  
the only non-zero contributions to the $P+Q_A$ projections are given by
\begin{equation}
\langle\Phi|(1+\Lambda_A)\overline{H}_{N,A} (P^{(N)}+Q_A^{(N)}) = 0 ~, \label{LambdaA}
\end{equation}
which represents $\Lambda$-equation for isolated sub-system A. Similar analysis leads to the form of (\ref{seeq2}) onto $Q_B$, 
\begin{equation}
\langle\Phi|\left(e^{S_B}\overline{H}_{N,B}\right)_c Q_B = 0 ~.
\label{SB}
\end{equation}
At this point, both (\ref{LambdaA}) and (\ref{SB}), as well as their $\Lambda_A$ and $S_B$ solutions, are independent of the $S_{AB}$ part. 

To determine $S_{AB}$ we  re-cast Eq. (\ref{seeq2}) in projections on 
$Q_{AB}$ in the form
\begin{widetext}
\begin{equation}
\langle\Phi|\lbrack \left(\left(e^{S_{AB}}-1\right)\overline{H}_{N,A}\right)_c + 
\left(\left(e^{S_{AB}}-1\right)\left(e^{S_B}\overline{H}_{N,B}\right)_c\right)_c \rbrack Q_{AB} = 0 ~,
\label{qabeq}
\end{equation}
\end{widetext}
which represents equations homogeneous in $S_{AB}$  with $S_{AB}=0$ as a solution. This proves that 
in the NSL 
\begin{equation} 
S_{\rm ext} \xrightarrow{NSL} S_B ~.
\label{SextNSL}
\end{equation}


\section{Properties of $X_K(\omega)$ and $Y_L(\omega)$ operators in the NSL} \label{appendix2}
In this appendix, we will discuss properties of $X_K(\omega)$ and $Y_L(\omega)$ operators in the NLS.
Although in previous papers \cite{kowalski16_062512} the connected character of $X_K(\omega)$ and $Y_L(\omega)$  operators has been discussed,
in this Appendix we will focus on explicit form of the equations for these operators in the NSL. For the sake of brevity in our analysis we focus on the $X_K(\omega)$. 
Results of this analysis can be readily extended to the $Y_L(\omega)$ operator. 
Same as Appendix \ref{appendix1}, we will use the asymptotic behavior of the following operators:
\begin{align}
&T \xrightarrow{NSL} T_A + T_B  ~, ~
T_{\rm int} \xrightarrow{NSL}  T_A ~,~
T_{\rm ext} \xrightarrow{NSL}  T_B ~.
\end{align}
Employ $K_A$ to denote the active spin-orbital $K$ in the NSL, 
the equations for $X_{K_A}(\omega)$ take the form 
\begin{widetext}
\begin{align}
Q^{(N-1)}_A \lbrack (\omega_{-i0}+\overline{H}_{N,A} +\overline{H}_{N,B} )  
(X_{K_A,A}(\omega) + X_{K_A,B}(\omega)+X_{K_{A},AB}(\omega)) \rbrack_c |\Phi\rangle &= 
Q^{(N-1)}_A\overline{a_{K_A}}|\Phi\rangle ~,\label{qae1} \\
Q^{(N-1)}_B \lbrack (\omega_{-i0}+\overline{H}_{N,A} +\overline{H}_{N,B})  
(X_{K_A,A}(\omega) + X_{K_A,B}(\omega)+X_{K_{A},AB}(\omega))\rbrack_c |\Phi\rangle &= 
Q^{(N-1)}_B\overline{a_{K_A}}|\Phi\rangle ~,\label{qae2} \\
Q^{(N-1)}_{AB}\lbrack  (\omega_{-i0}+\overline{H}_{N,A} +\overline{H}_{N,B})  
(X_{K_A,A}(\omega) + X_{K_A,B}(\omega)+X_{K_{A},AB}(\omega)) \rbrack_c |\Phi\rangle &= 
Q^{(N-1)}_{AB}\overline{a_{K_A}}|\Phi\rangle ~,\label{qae3}
\end{align}
\end{widetext}
where in analogy to analysis in Appendix \ref{appendix1}
\begin{align}
X_{K,{\rm int}}(\omega) &\xrightarrow{NSL} X_{K_A,A} (\omega)~, \label{XintNSL} \\
X_{K,{\rm ext}}(\omega) &\xrightarrow{NSL}  X_{K_A,B}(\omega)+X_{K_A,AB}(\omega) ~. \end{align}
Here $X_{K_A,A} (\omega)$, $X_{K_A,B}(\omega)$, and $X_{K_A,AB}(\omega)$ are the operators defined by the excitations localized on sub-system A, B, and the excitations that involve both sub-systems simultaneously, respectively.   The operators  $Q_A^{(N-1)}$, $Q_B^{(N-1)}$, and $Q_{AB}^{(N-1)}$ are defined in an analogous way in the $(N-1)$-electron Hilbert space. 
Given the fact that creation/annihilation operators corresponding to 
electronic states on different sub-systems anti-commute, one can write that 
\begin{equation}
\overline{a_{K_A}}=e^{-T_A} a_{K_A} e^{T_A} ~,
\label{apa}
\end{equation}
and the fact that only connected expressions enter $X_{K_A}(\omega)$ equations, the Eqs. (\ref{qae1}-\ref{qae3}) can be simplified to 
\begin{widetext}
\begin{align}
Q^{(N-1)}_A \lbrack (\omega_{-i0}+\overline{H}_{N,A}) X_{K_A,A}(\omega)+\overline{H}_{N,B} X_{K_A,AB}(\omega)  ) \rbrack_c |\Phi\rangle &= 
Q^{(N-1)}_A\overline{a_{K_A}}|\Phi\rangle ~,\label{qae1n} \\
Q^{(N-1)}_B \lbrack( \omega_{-i0}+\overline{H}_{N,B}) X_{K_A,B}(\omega)+(\overline{H}_{N,A} +\overline{H}_{N,B} )X_{K_{A},AB}(\omega)
\rbrack_c |\Phi\rangle &= 0 ~,\label{qae2n} \\
Q^{(N-1)}_{AB}\lbrack  (\omega_{-i0}+\overline{H}_{N,A}+\overline{H}_{N,B}) X_{K_A,AB}(\omega)
 \rbrack_c |\Phi\rangle &= 0  ~.\label{qae3n}
\end{align}
\end{widetext}
A close inspection of Eqs. (\ref{qae2n}-\ref{qae3n}) shows that both $X_{K_A,AB}(\omega)$ and $X_{K_A,B}(\omega)$
vanish, i.e.,
\begin{align}
X_{K_{A},AB}(\omega) &= 0 ~, \label{zero1} \\
X_{K_A,B}(\omega) &= 0 ~, \label{zero2} 
\end{align}
and thus
\begin{align}
X_{K,\rm ext} \xrightarrow[]{NSL} 0 ~,~
X_{K} \xrightarrow[]{NSL} X_{K_A,A} \label{XNSL}
\end{align}
for non-resonant frequencies $\omega$. This implies that the equation for $X_{K_A,A}(\omega)$ takes the form 
\begin{equation}
Q^{(N-1)}_A \lbrack (\omega_{-i0}+\overline{H}_{N,A}) X_{K_A,A}(\omega)  \rbrack_c |\Phi\rangle =
Q^{(N-1)}_A\overline{a_{K_A}}|\Phi\rangle ~,\label{xafinal} 
\end{equation}
and is identical to the equation for $X_{K_A}(\omega)$'s of the isolated sub-system A. 
Analogous results can be derived for the $Y_L(\omega)$ operator, which in the NSL is localized only on sub-system A, i.e.,
\begin{align}
Y_L(\omega) \xrightarrow{NSL} Y_{L_A,A}(\omega), \label{YNSL}    
\end{align}
where $Y_{L_A,A}(\omega)$ operator satisfies the equations: 
\begin{equation}
Q^{(N+1)}_A \lbrack (\omega_{+i0} - \overline{H}_{N,A}) Y_{L_A,A}(\omega) \rbrack_c |\Phi\rangle =
Q^{(N+1)}_A\overline{a^{\dagger}_{L_A}}|\Phi\rangle ~,\label{yafinal} 
\end{equation}
with $\overline{a^{\dagger}_{L_A}}=e^{-T_A} a^{\dagger}_{L_A}e^{T_A}$. This analysis also illustrates the size-intensivity of 
the $X_K(\omega)$ and $Y_L(\omega)$ operators.


\section{Asymptotic properties of sub-systems CC Green's function} \label{appendix3}
Using properties of the $T_{\rm int}$, $T_{\rm ext}$, $\Lambda_{\rm int}$, $S_{\rm ext}$,  $X_K(\omega)$, 
and $Y_L(\omega)$ operators derived in the Appendices \ref{appendix1} and \ref{appendix2}, we can derive properties of the sub-block of the 
CC Green's function matrix. 
Following Eqs. (\ref{TNSL},\ref{SextNSL}),  (\ref{XNSL},\ref{YNSL}),  one can derive the form of 
$G_{KL}(\omega)$ block in the NSL as: 
\begin{align}
G_{KL}(\omega) \xrightarrow{NSL}&~~~ \langle\Phi|(1+\Lambda_A)\overline{a^{\dagger}_{L_A,A}} X_{K_A,A}(\omega)|\Phi\rangle \notag \\
& +\langle\Phi|(1+\Lambda_A)\overline{a_{K_A,A}} Y_{L_A,A}(\omega)|\Phi\rangle ~,
\label{gnsl}
\end{align}
which is identical with the algebraic form of the CC Green's function for isolated sub-system A.

Similarly, for other blocks in the CC Green's function matrix, we have
\begin{align}
G_{\overline{p}\overline{q}}(\omega) \xrightarrow{NSL}&~~~ \langle \Phi | (1+\Lambda_B) \overline{a^{\dagger}_{\overline{q}_B}}_{,B} X_{\overline{p},B}(\omega)| \Phi \rangle \notag \\
&+\langle \Phi | (1+\Lambda_B) \overline{a_{\overline{p}_B}}_{,B} Y_{\overline{q},B}(\omega)| \Phi \rangle ~, \label{gnsl} \\
G_{K\overline{p}}(\omega) \xrightarrow{NSL}& ~~0 ~,~
G_{\overline{p}K}(\omega) \xrightarrow{NSL} ~0.
\end{align}
Therefore, in the NSL the $\mathbf{G}$ of the composite system $\textbf{AB}$ simply becomes a diagonal block matrix
\begin{align}
    \mathbf{G}^{\rm \textbf{AB}}(\omega) \xrightarrow{NSL}  
    \left[\begin{array}{c|c} \mathbf{G}^{\rm \textbf{A}}(\omega) & \textbf{0}\\ \hline
    \textbf{0} & \mathbf{G}^{\rm \textbf{B}}(\omega) \end{array} \right].
\end{align}
%


\bibliography{gfcc}

\end{document}